\documentclass[%
reprint,
 amsmath,amssymb,
aps,
prb,
]{revtex4-1}

\usepackage[final]{graphicx}
\usepackage{dcolumn}
\usepackage{float}
\usepackage{caption}
\usepackage{amsmath}
\usepackage{csquotes}
\usepackage{subfig}
\usepackage{blindtext}

\usepackage{changepage}
\usepackage{physics}
\usepackage{color,soul}
\usepackage[dvipsnames]{xcolor}
\usepackage{bm}
\usepackage{mathtools}

\begin{document}

\preprint{APS/123-QED}

\title{Deformed Morse-like potential}

\author{I. A. Assi}
\email{iassi@mun.ca}%
\affiliation{Department of Physics and Physical Oceanography, Memorial University of Newfoundland, St. John’s, Newfoundland \& Labrador, A1B 3X7, Canada}

\author{A. D. Alhaidari}
\affiliation{Saudi Center for Theoretical Physics, P.O. Box 32741, Jeddah 21438, Saudi Arabia}

\author{H. Bahlouli}
\affiliation{Physics Department, King Fahd University of Petroleum  Minerals, Dhahran 31261, Saudi Arabia}

\date{\today}
\begin{abstract}
We introduce an exactly solvable one-dimensional potential that supports both bound and/or resonance states. This potential is a generalization of the well-known 1D Morse potential where we introduced a deformation that preserves the finite spectrum property. On the other hand, in the limit of zero deformation, the potential reduces to the exponentially confining potential well introduced recently by A. D. Alhaidari. The latter potential supports infinite spectrum which means that the zero deformation limit is a critical point where our system will transition from the finite spectrum limit to the infinite spectrum limit. We solve the corresponding Schrodinger equation and obtain the energy spectrum and the eigenstates using the tridiagonal representation approach.
\end{abstract}

\maketitle
\section{Introduction}
Studying exactly solvable potentials has been an issue of great interest since the birth of quantum mechanics. Many techniques have been designed aiming at obtaining exact solution to the Schrodinger equation. These methods include, but not limited to, factorization method, point canonical transformation, supersymmetry approach, shape invariance, Darboux transformation, second quantization, asymptotic iteration method, group theoretical approaches, path integral transformation and Nikiforov-Uvarov method. For a brief  description of these methods the reader can consult the corresponding literature.\cite{COOPER,Bander,ALHASSID1,Infeld,Ciftci,De,NU} Our group has devised a new algebraic 
method called the “Tridiagonal Representation Approach”(TRA) which enabled  us to enlarge the class of exactly solvable potentials. For more details on the TRA the reader is encouraged to refer to the recent summary of this method and references therein.\cite{TRA2019}  In this approach the exact solution of the corresponding Schrodinger equation is expressed in the form of a bounded series of suitably selected square integrable basis set. Requiring a tridiagonal representation of the wave equation in this basis set will result in a three  term recursion relation for the expansion coefficients of the series which are then solved in terms  orthogonal polynomials in the energy and physical parameter spaces.

In the present work we introduce the following exactly solvable 1D deformed Morse-like potential 
\begin{equation}
\label{eqn:potentialVq}
   {{V}_{q}}(x)=A{{\left({{e}^{\lambda x}}+q\right)}^{-2}}+B\left( {{e}^{\lambda x}}+q\right)^{-1}+C{{e}^{\lambda x}}-\frac{A+qB}{q^2}
\end{equation}
where $x\in (-\infty, +\infty)$ and the potential parameters $\{q, A, B, V, \lambda\}$ are real such that $q>0$, $C>0$  and $q$ is the deformation parameter. 
The design of the above potential is such that it reduces to zero at minus infinity (the last constant value in (\ref{eqn:potentialVq}) ensures that) and blows up at positive infinity. With these choices the potential will support both resonances and/or bound states or none depending on the range of potential parameters. For intense, for $C=q=0$, this potential reduces to the well-known 1D Morse potential with a finite energy spectrum,\cite{MorseI} and when $q=0$, this becomes the exponentially confining potential well introduced recently by Alhaidari with an infinite energy spectrum.\cite{Alhaidariexpon} For other values of the potential parameters we will see that we can have different situations where only resonances, only bound states, both bound and resonant states or none are allowed. Actually, we were able to generate an illuminating spectral phase diagram that shows different regions of the potential parameter space where bound states and/or resonances can occur.

The rest of this manuscript is organized as follows: In section II we investigate the analytical properties of the deformed Morse-like potential (\ref{eqn:potentialVq}). In section III we present the theoretical derivation of the solvable potential (\ref{eqn:potentialVq}) using the tridiagonal representation approach. In section IV we compute the energy spectrum associated with this potential using the potential parameter spectrum (PPS),\cite{PPSI} which is a highly non traditional eigenvalue problem. In section IV we confirm the validity of our generated energy spectrum using a direct numerical diagonalization technique in a suitable $L^2$ basis, and the asymptotic iteration method.\cite{Ciftci,Sous} In section VII we present a conclusion and discuss possible extensions of this work and its potential application.
\section{The potential Structure}
\begin{figure*}
\subfloat[]{\includegraphics[height=2.0in,width = 2.0in]{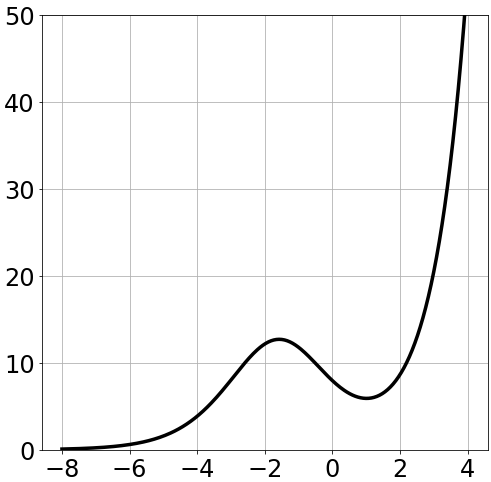}}\hfill
\subfloat[]{\includegraphics[height=2.0in,width = 2.0in]{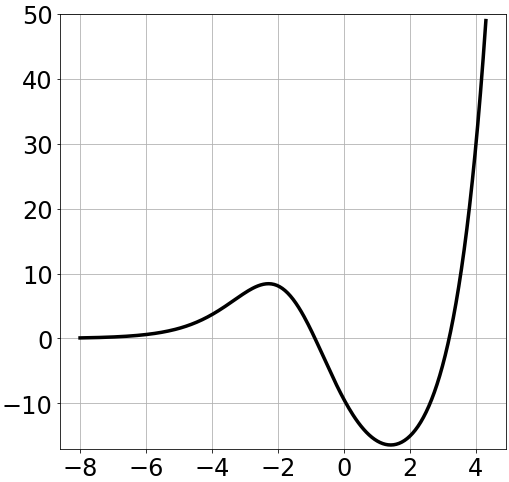}}\hfill
\subfloat[]{\includegraphics[height=2.0in,width = 2.0in]{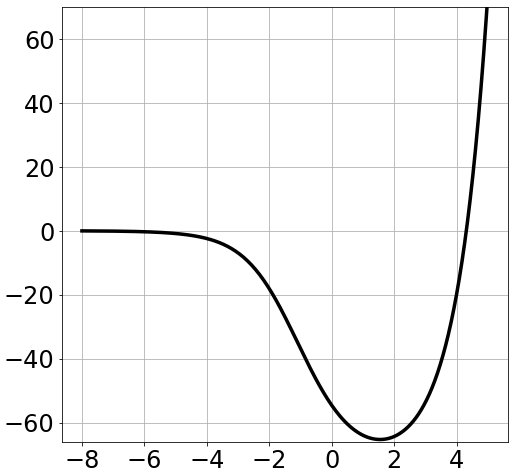}}\hfill
\subfloat[]{\includegraphics[height=2.0in,width = 2.0in]{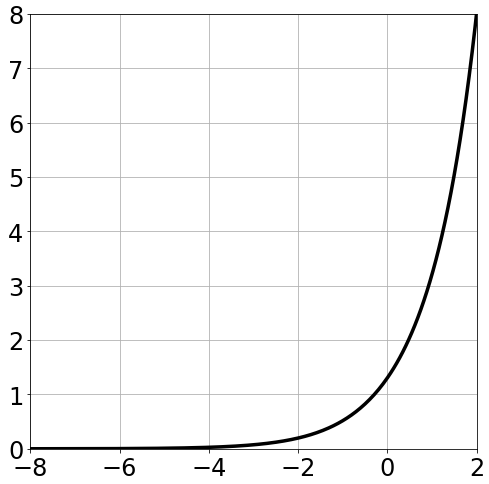}}\hspace{0.1\textwidth}
\subfloat[]{\includegraphics[height=2.0in,width = 2.0in]{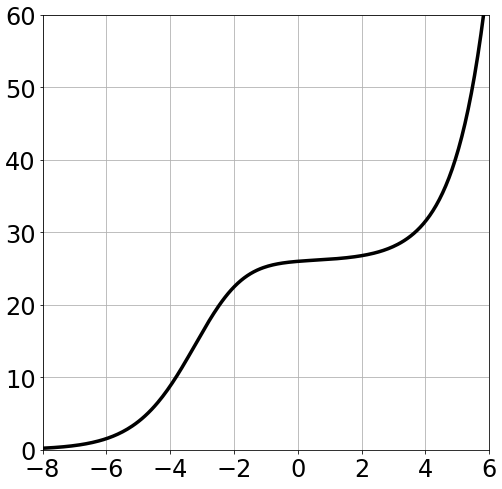}}
\captionsetup{justification   = raggedright,
              singlelinecheck = false}
\caption{Different possible configurations for the potential given by Eq. \ref{eqn:potentialVq} which  might support (a) only resonances, (b) bound and resonance states , (c) only bound states, and (d) \& (e) none. All configurations support scattering states.}
\label{fig:pot}
\end{figure*}

In this section we analyze all possible potential configurations that can sustain bound and/or resonance states. Starting with the potential (\ref{eqn:potentialVq}), we first look at possible extrema which are given by the root of the equation $\frac{dV_q}{dx}=0$ at $x=x_0$, we then get
\begin{equation}
\label{eqn:cubiceq}
    2A z_0^3 + B z_0^2-C=0
\end{equation}
where $z_0^{-1}={{e}^{\lambda x_0}}+q$. If $A>0$ and $B<0$, then Descartes' rule of signs dictates that this cubic equation has a single real positive root.\cite{signsrule} Now, if $0 < z_0 < 1/q$, then the potential has a minimum and it can support only bound states. The second scenario holds when $B>0$ and $A<0$, then Descartes' sign rule guarantees that this cubic equation has either two real positive root or none. Now, if both roots pertained to the interval $(0,1/q)$ and have unequal values then the potential has a minimum and a maximum and it can then support either resonances or a mix of resonance and bound states. Thirdly, if $A>0$ and $B<0$, then Descartes' sign rule implies that the cubic equation (\ref{eqn:cubiceq}) has only one positive real root, similarly to the first scenario, this case can support only bound states. Lastly, when $A<0$ and $B<0$, we have no real positive root and the potential cannot support neither bound states nor resonances. Obviously, in all cases, the potential supports scattering states. These different situations are summarized in five self explanatory panels in Fig.\ref{fig:pot}. We have also created a video animation of the potential that shows the continuous transition from figures \ref{fig:pot}a to \ref{fig:pot}b as the potential parameters A/C and B/C are varied continuously. 

\begin{figure*}
\subfloat[]{\includegraphics[width = 3.55in]{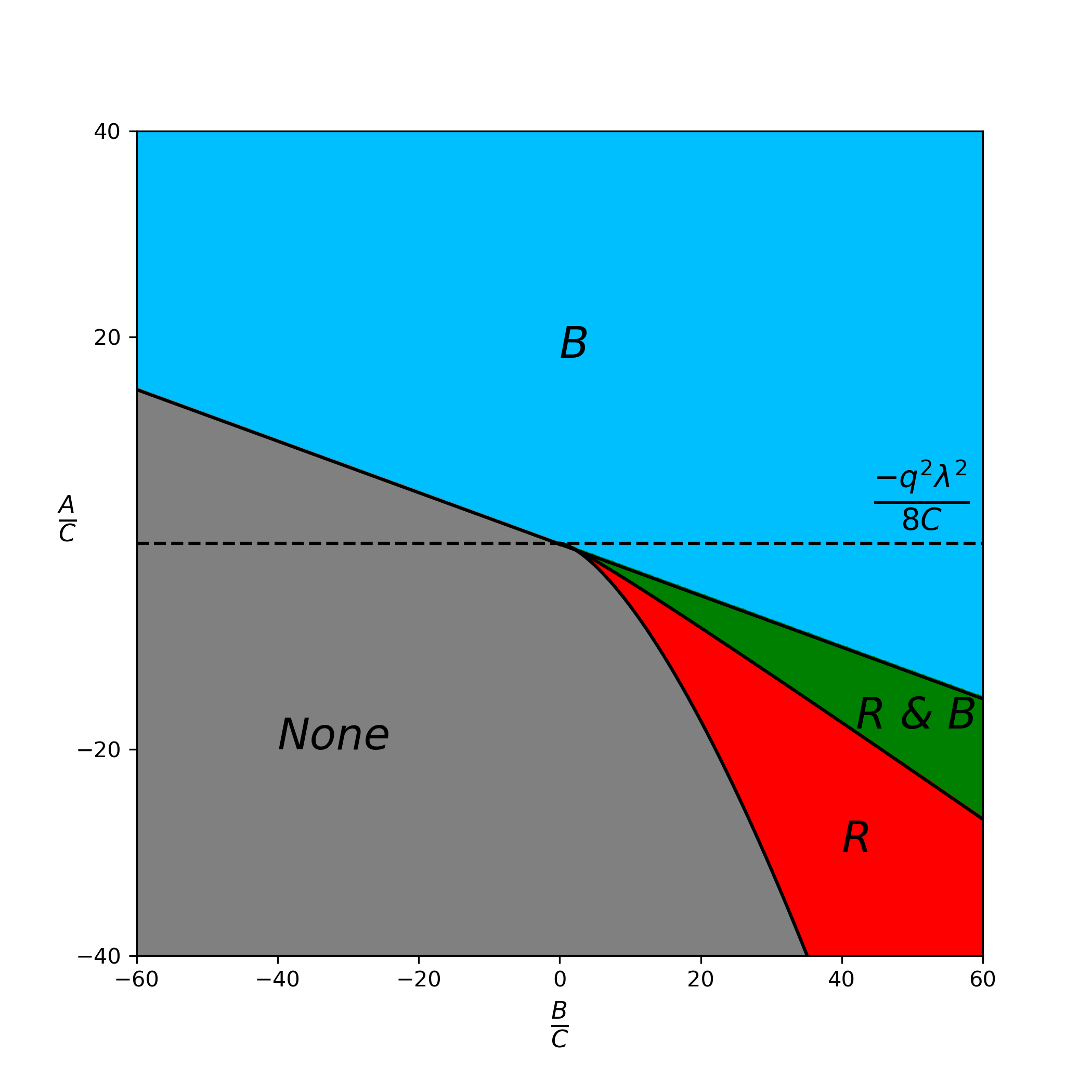}}
\subfloat[]{\includegraphics[width = 3.45in]{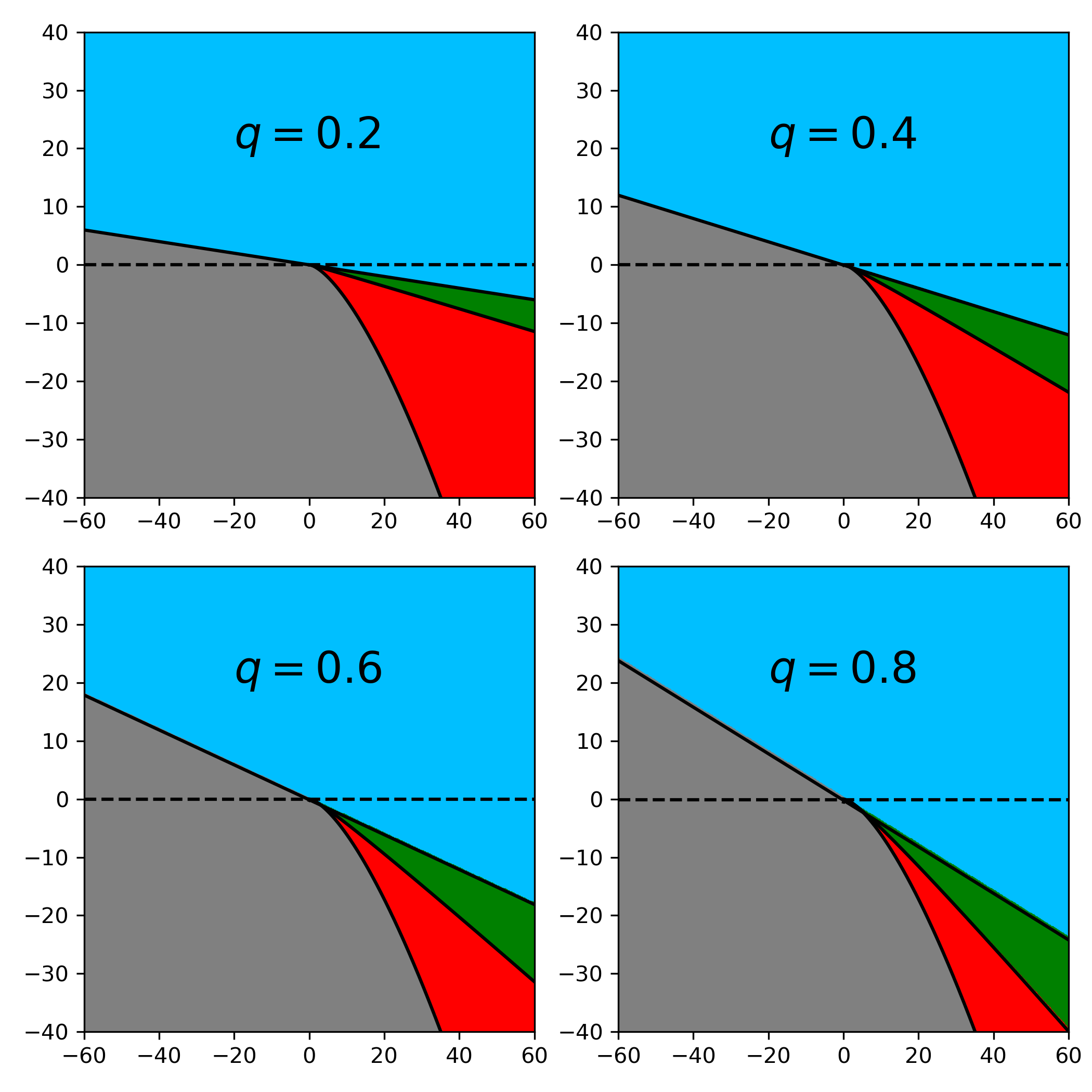}}
\captionsetup{justification   = raggedright,
              singlelinecheck = false}
\caption{(a) Snapshot of a video animation of the spectral phase diagram showing several regions where only bound (B) states can exist in the blue region, only resonances (R) in the red region, a mixture of resonance and bound states in the green region, and the gray region where none of these states can exist. The dashed line represents the TRA lower limit on the potential parameter $A$ scaled in units of $C$. Here we took $\lambda=1$ and $q=0.5$, and in (b) the spectral phase diagram for several values of $q$, from top left to bottom right we took $q=0.2,0.4,0.6,$ and $0.8$.}
\label{fig:phaseS}
\end{figure*}

On the other hand, we plotted the \textit{spectral phase diagram} in Fig.\ref{fig:phaseS} which shows different physical situations that the potential (\ref{eqn:potentialVq}) can support based on the values of the parameters $\{A,B\}$ which are scaled in units of $C$. In Fig. \ref{fig:phaseS}(a), we took $q=0.5$ and the scaling parameter $\lambda$ is set to unity. The blue region represents part of the spectral phase diagram (SPD) where only bound (B) states can exist. In the green region, a mix of bound (B) and resonance (R) states whereas in the red region, only resonances. The fourth case is when neither bound nor resonance states can occur in the grey region. On the other hand, in Fig. \ref{fig:phaseS} (b), we varied $q$ and plotted the SPD as indicated. We find that increasing $q$ from $0.2$ to $0.8$ has resulted in increasing the size of the blue and green regions and suppressed the red region while the boarder between red and grey regions remains fixed (independent of $q$). Thus, the deformation parameter $q$ can change the state of the system at fixed parameters $\{A,B,C\}$, such as transforming regions from bearing resonances only to allowing a mix of resonance and bound states. We have also provided a video animation showing how the SPD changes with $q$ which we varied from $0.01$ to $0.8$ with equal steps of $0.01$. 

Finally, we find the equations of the boundaries in the SPD as follows. The boundary of the blue region is obtained by setting $V_q(x)=V'_q(x)=0$ and $z_0=1/q$, giving 
\begin{equation}
\label{eqn:BGboundary}
    2A=q^3C-qB
\end{equation}
the boundary between the grey region and the red region is obtained by solving $V'_q(x)=V''_q(x)=0$, 
\begin{equation}
\label{eqn:RGboundary}
    B^3=27A^2C
\end{equation}
with $A<0$, $B>0$, and is independent of $q$. The third boundary is between the green and red regions and is obtained by solving $V_q(x)=V'_q(x)=0$ allowing only one of the roots $z_0$ to be different from $1/q$, we obtain
\begin{equation}
\label{eqn:RGreenBoundary}
    A=q^2\sqrt{BC}-q\left(B+q^2C\right)
\end{equation}

\section{TRA Formulation} 
Starting with the 1D Schrodinger equation (in units of $\hbar = m = 1$),
\begin{equation}
   \left[-\frac{1}{2}\frac{d^2}{dx^2}+V(x)-E\right]\psi(x)=0 
\end{equation}
we make a change of variable $y=\frac{2}{q}e^{\lambda x}+1\geq 1$, giving 
\begin{equation}
\label{eqn:SEinY}
    \left[ {{\left( y-1 \right)}^{2}}\frac{{{d}^{2}}}{d{{y}^{2}}}+\left( y-1 \right)\frac{d}{dy}+\varepsilon -U(y) \right]\psi \left( y \right)=0
\end{equation}
where $\varepsilon=2E/\lambda^2$ and $U(y)=2V(y)/\lambda^2$. Following the TRA formalism, we expand the wavefunction as $\psi(y)=\sum_{n}f_{n}(E,\mathcal{P})\phi_{n}(y)$, where the expansion coefficients $f_{n}(E,\mathcal{P})$ are generally functions of the energy and potential parameters that are lumped together in $\mathcal{P}$. In our case $\mathcal{P}=\{A,B,C,q\}$, and the suitable square integrable basis function $\phi_{n}(y)$ is \cite{TRA2019}
\begin{equation}
\label{eqn:basisTRA}
    {{\phi }_{n}}(x)={{A}_{n}}{{\left( y-1 \right)}^{\frac{\mu}{2}}}{{\left( y+1 \right)}^{\frac{\nu+1}{2} }}P_{n}^{(\mu ,\,\nu )}(y)
\end{equation}
where $P_{n}^{(\mu ,\,\nu )}(y)$ is the finite Jacobi polynomial defined in Appendix \ref{app:Jacobi}, $n=0,1,\cdots,N$ for some non-negative integer $N$, $A_{n}$ is a normalization constant defined in Eq.\ref{eqn:An}, and $\{\mu,\nu\}$ are real basis parameters with $\mu>-1$ and $\mu+\nu<-2N-1$.

Using the differential equation of the finite Jacobi polynomials Eq. \ref{eqn:DffJac}, the action of the operator in Eq. \ref{eqn:SEinY} on the basis Eq.\ref{eqn:basisTRA} is given by 

\begin{widetext}
\begin{multline}
\left[(y-1)^{2}\frac{d^2}{dy^2}+(y-1)\frac{d}{dy}+\varepsilon-U(y)\right]\phi_n(y)=A_n (y-1)^{\frac{\mu}{2}+1}(y+1)^{\frac{\nu-1}{2}}\Bigg\{n(n+\mu+\nu+1)+\frac{\mu^2}{4}\frac{y+1}{y-1}\\
+\frac{\nu^2-1}{4}\frac{y-1}{y+1}+\frac{(\mu+1)(\nu+1)}{2}+(\varepsilon-U(y))\frac{y+1}{y-1}\Bigg\}P_{n}^{(\mu ,\,\nu )}(y)
\end{multline}
\end{widetext}
Requiring that the representation of this operator in the basis set $\phi_{n}(x)$  be tridiagonal and looking for energy independent potential solutions we end up with the requirements   $\varepsilon=-\frac{\mu^2}{4}$, and
\begin{equation}
    \frac{1}{4}\left( {{\nu }^{2}}-1 \right)\frac{y-1}{y+1}-U(y)\frac{y+1}{y-1}=-Fy-D
\end{equation}
This last requirement gives rise to the solvable deformed Morse-like potential announced at the beginning of this manuscript (1) where the parameters $\{\nu,F,D\}$ are related to $\{A,B,C,q,\lambda\}$ as indicated in Table.\ref{tab:parametersmap}. This potential by construction gives rise to the tridiagonal representation of the wave operator $(\hat{H}-E)$ in this basis,
\begin{equation}
\label{eqn:TRArepL}
    -\frac{2}{{{\lambda }^{2}}}\left\langle m|(H-E)\left. |n \right\rangle  \right.=a_n{{\delta }_{n,m}}-F\langle m|y\left. |n \right\rangle 
\end{equation}
where we defined $\langle m|(...)|n \rangle:=\lambda\int_{-\infty}^{+\infty}dx \phi_m(x) (...)\phi_n(x)$, $a_n=n(n+\mu +\nu +1)+\frac{1}{2}\left( \mu +1 \right)\left( \nu +1 \right)-D$, and $\langle m|y|n\rangle$ is the tridiagonal matrix defined in Eq.\ref{eqn:overlap}.\\

\begin{table}[h!]
  \begin{center}
    \caption{The relation between the parameters of $U(x)$ in terms of those of $V_q(x)$.}
    \label{tab:parametersmap}
    \begin{tabular}{|c|c|} 
    
      \hline
      $U(x)$ & $V_q(x)$\\
      \hline
$\nu$&$-\sqrt{\frac{8A}{\lambda^2q^2}+1}$\\
$D$&$\frac{qC}{\lambda^2}-\frac{2B}{q\lambda^2}-\frac{4A}{q^2\lambda^2}$\\
$F$&$\frac{qC}{\lambda^2}$\\
        \hline
    \end{tabular}
  \end{center}
\end{table}

It is clear that the above representation is diagonal when $F=0$ (i.e. $C=0$), resulting in the following spectrum\cite{RMI,RMII} 
\begin{equation}
    2{{E}_{n}}/{{\lambda }^{2}}=-\frac{1}{4}{{\left\{ \frac{D+\frac{1}{4}\left( {{\nu }^{2}}-1 \right)}{\left[ n+\frac{1}{2}\left( \nu +1 \right) \right]}-\left[ n+\frac{1}{2}\left( \nu +1 \right) \right] \right\}}^{2}}
\end{equation}
and associated wavefunction, 
\begin{equation}
\label{eqn:diagWF}
    {{\psi }_{n}}(x)=\mathcal{A}_{n}{{\left( y-1 \right)}^{\mu_n/2 }}{{\left( y+1 \right)}^{\frac{\nu+1}{2} }}P_{n}^{(\mu_n ,\,\nu )}(y)
\end{equation}
where $\mathcal{A}_{n}$ is a normalization constant.\\

To find the solution when $F\neq 0$ (i.e. $C > 0$), we use Eq.\ref{eqn:TRArepL} to express the expansion coefficients of the wavefunction $\{f_n(E,\mathcal{P})\}$ in terms of a three-term recursion relation,
\begin{equation}
    \left[FQ_n+a_n\right]{{f}_{n}}(E )=F\left[ {{S}_{n-1}}{{f}_{n-1}}(E )+{{S}_{n}}{{f}_{n+1}}(E) \right]
\end{equation}
where $\{Q_n,S_n\}$ are defined in Eqs.\ref{eqn:C} \& \ref{eqnD}. Making the substitution $f_{n}=\frac{A_nf_0\xi_{n}}{A_0}$, we obtain the following three-term recursion relation for $\xi_{n}$,
\begin{eqnarray}
\frac{1}{F}a_{n}{{\xi}_{n}}(E)=\tfrac{2(n+\mu )(n+\nu )}{\left( 2n+\mu +\nu  \right)\left( 2n+\mu +\nu +1 \right)}{{\xi}_{n-1}}(E)
\nonumber\\
+\tfrac{2\left(n+1 \right)(n+\mu +\nu +1)}{(2n+\mu +\nu +1)\left( 2n+\mu +\nu +2 \right)}{{\xi}_{n+1}}(E)-{{Q}_{n}}{{\xi}_{n}}(E)
\end{eqnarray}
we deduce that $\xi_{n}=\bar{H}_n^{(\mu,\nu)}(-F^{-1};\ell,\pi/2)$, where $\bar{H}^{\mu,\nu}(z^{-1};\ell,\theta)$ is an orthogonal polynomial introduced recently by Alhaidari,\cite{Assche} and $\ell=\frac{1}{2}\left( \mu +1 \right)\left( \nu +1 \right)-{{\left( \frac{\mu +\nu +1}{2} \right)}^{2}}-D$. This polynomial is defined by its three term recursion relation and some of its properties were addressed by W. Van Assche.\cite{Assche} However, we do not know its important analytical properties such as its weight function, generating function, asymptotics, orthogonality relations, zeros, $\dots$ etc. We hope that experts in the field of orthogonal polynomials will be able to study this polynomial and derive its analytical properties.\\ 

In addition, the analytical properties of this polynomial will also be useful when considering the exponentially confining potential well problem (Ref.\onlinecite{Alhaidariexpon}) obtained from our potential (1) by considering the zero deformation limit. Although this system supports an infinite spectrum and the Jacobi basis used in this work is finite, in the limit $q \to 0$ the Jacobi basis becomes infinite! This can be seen through $\nu=-\sqrt{\frac{8A}{\lambda^2q^2}+1}$ along with $N<-\left(\mu+\nu+1\right)/2$. This is also in agreement with the fact that the polynomial $\bar{H}^{\mu,\nu}(z^{-1};\ell,\theta)$ has discrete infinite spectrum.\cite{Assche} 

Consequently, the wavefunction associated with the potential (\ref{eqn:potentialVq}) is
\begin{equation*}
    \psi_n(x)=R_n(y-1)^{\frac{\mu}{2}}(y+1)^{\frac{\nu+1}{2}}\times
\end{equation*}
\begin{equation}
    \sum_{k=0}^{N}c_k\bar{H}_k^{(\mu,\nu)}(-F^{-1};\ell,\pi/2)P_{k}^{(\mu ,\,\nu )}(y)
\end{equation}
where $R_n$ is some normalization constant. 

We should mention that the TRA imposes the constraint $A\geq - \lambda^2q^2/8$ as per the results in Table.\ref{tab:parametersmap}. The critical limit $A=- \lambda^2q^2/8$ is represented by the dashed lines in Fig.\ref{fig:phaseS}. It is clear that at some values of $q$, the blue triangle below this critical limit may contain bound states that cannot be accounted for by our TRA solution but can be treated together with resonance states by other numerical means.\cite{complexs1}      

\section{Potential Parameter Spectrum} 
The potential parameter spectrum (PPS) is a non traditional numerical eigenvalue method that originated in the context of the TRA whenever the basis set is energy dependent \cite{PPSI}. This happens, for example, when the TRA constraints dictate that one or more of the basis parameters is/are energy dependent. Such is the case in our current problem where the basis parameter $\mu$ is required to be equal to $2\sqrt{-2E/\lambda^2}$. Below, we show how to obtain the TRA solution (energy spectrum and wavefunction) in this case.

The fundamental TRA equation in the basis $\{{{\phi }_{n}}(x)\}$ is

\begin{equation}
\label{eqn:pps1}
    \hat{J}{{\phi }_{n}}(x)=\omega(x)\left[c_n{{\phi }_{n}}(x)+b_{n-1}{{\phi }_{n-1}}(x)+b_n{{\phi }_{n+1}}(x)\right]
\end{equation}
where $\hat{J}=\hat{H}-E$ is the wave operator, $\omega(x)$ is a node-less entire function, and $\{b_n,c_n\}$ are real coefficients. Now, we write $c_n=a_n-z$ where $z$ is some proper parameter such that $\{a_n,b_n\}$ are independent of $z$. Then, writing the wavefunction as series  $\psi(x)=\sum_{n}f_{n}\phi_{n}(x)$ with $f_n=f_0P_n(z)$, then the wave equation $\hat{J}\psi(x)=0$ gives the following three-terms recursion relation 
\begin{equation}
\label{eqn:pps2}
    zP_{n}(z)=a_{n}P_{n}(z)+b_{n-1}P_{n-1}(z)+b_{n}P_{n+1}(z)
\end{equation}
Favards theorem asserts that the solution of such a three term recursion relation is a set of orthogonal polynomials of degree $n$ in the variable $z$, $\{P_{n}(z)\}$. These polynomials can be computed to any desirable degree using the above three term recursion relation along with the initial conditions. However, the full analytical properties will be known only if the associated weight function, generating function and asymptotic limits are known. 

In our current situation, however, we have two difficulties: (1) the analytic properties of $P_n(z)$ are not known, and (2) the basis and $\{a_n, b_n\}$ are energy dependent. Under these circumstances we resort to an indirect numerical approach, the potential parameter spectrum (PPS), to evaluate the corresponding energy spectrum. In brief the main components of the PPS approach go as follows:
\begin{enumerate}
    \item The polynomial argument $z$ must be chosen such that it does not depend on the energy
and contains at least one of the system’s parameters that does not appear in $\{a_n, b_n\}$. Let us call that parameter $\gamma$ and write $z$ as $z(\gamma)$.
\item Write Eq. \ref{eqn:pps2} as $z(\gamma)\ket{P}=T\ket{P}$ and take the tridiagonal matrix $T$ to be of finite size $N\times N$.
\item We choose a value for $E$ from a proper range and calculate $\{a_n, b_n\}_{n=0}^{N-1}$.
\item Calculate the eigenvalues of $z(\gamma)\ket{P}=T\ket{P}$ as $\{z_n(\gamma)\}_{n=0}^{N-1}$.
\item Repeat steps 3 and 4 for another $E$ in the chosen range until the whole range is covered.
\item Let us designate $\{\gamma_n^k\}_{n=0}^{N-1}$ as the resulting eigenvalues from the $k$th step that corresponds to the energy $E_k$. This set is called the "potential parameter spectrum" for $\gamma$ at energy $E_k$.
\item Sort the set $\{\gamma_n^k\}_{k=0}^{N-1}$ and for each fixed $n$, make a function fit of $\{E_k\}_{k=1}^{k=M}$ versus $\{\gamma_n^k\}_{k=0}^{N-1}$ and call this function $G(n,\gamma)$.
\item Thus, the energy spectrum of the system corresponding to a given physical parameter $\gamma$
is $\{G(m,\gamma)\}_{m=0}^{M}$, where $M$ is the maximum integer such that $G(m,\gamma)<0$.
\end{enumerate}

Now, the wavefunction $\psi_{m}(x)$ at the energy eigenvalue $G(m,\gamma)$ is obtained as follows. First, Calculate $\{a_n, b_n\}_{n=0}^{N-1}$ at the energy $G(m,\gamma)$ and then solve Eq.\ref{eqn:pps2} for $P_n(z(\gamma))$. Consequently, we obtain $\psi_m(x)=f_0(z)\sum_nP_n(z(\gamma))\phi_n(x)$ where the energy eigenvalue $G(m,\gamma)$ is implicit in the basis and energy polynomial.

\section{Numerical Diagonalization} 

For completeness and to give an independent verification of our results, we choose a numerical diagonalization approach that falls within the spirit of TRA. We start by writing our original Hamiltonian in the following form, 
\begin{equation}
    \hat{H}=\hat{H}_{0}+U_{q}(x)
\end{equation}
where $\hat{H}_{0}=-\frac{1}{2}\frac{d^2}{dx^2}+Ce^{\lambda x}$ and $U_{q}(x)=V_{q}(x)-Ce^{\lambda x}$. This suitable selection of $\hat{H}_{0}$ will enable us to treat exactly the $e^{\lambda x}$ term and leave only the short range part of the potential for numerical approximation. Then we select a basis set in which $\hat{H}_{0}$ can be represented by a tridiagonal matrix. In our case, the Laguerre basis will do the job,
\begin{equation}
\phi_n(x) = A_ny^{\frac{\gamma+1}{2}} e^{-y/2}L_n^\gamma(y),
\label{eq:4-1}
\end{equation}
where $A_n = \sqrt{n!/\Gamma(n+\gamma+1)}$ and $L_n^\gamma(y)$ is the associated Laguerre polynomial with $\gamma>-1$, $y=\rho e^{\lambda x/2}$, where $\rho=4\sqrt{2C}/\lambda$, giving  
\begin{equation}
   16\lambda^{-2}(\hat{H}_{0})_{n,m}=\alpha_n\delta_{n,m}-\beta_{n}\delta_{n,m+1}- \beta_{n+1}\delta_{n,m-1} 
\end{equation}
where 
\begin{equation}
    \alpha_n=\left( 2n + \gamma+ 1 \right)^{2}-\frac{1}{2}\left(\gamma^2-1\right),
\end{equation}
\begin{equation}
    \beta_{n}=\left(2n+\gamma\right)\sqrt{n(n+\gamma)}.
\end{equation}
Thus in the above basis set the matrix element of the seed Hamitonian $\hat{H}_{0}$ can be computed exactly, however, the potential terms $U_q(x)$ in this basis is calculated using the numerical Gauss quadrature approach
\begin{equation}
    U_{q}=\Lambda.D.\Lambda^{T},
\end{equation}
where $\Lambda$ is the normalized eigenvector matrix associated with the following quadrature matrix
\begin{equation}
   \langle m|y|n\rangle=c_n\delta_{n,m}-d_n\delta_{n,m+1}-d_{n+1}\delta_{n,m-1}
\end{equation}
where $\langle m|y|n\rangle=A^{2}_n\int_0^{\infty}dx x^{\gamma+1} e^{-x} L^{\gamma}_m(x)L^{\gamma}_n(x)$, $c_n=2n + \gamma + 1$, $d_n=\sqrt{n(n+\gamma)}$, and
\begin{equation}
    D_{ij}=\left[\frac{A}{\left[(e_i/\rho)^2+q\right]^2}+\frac{B}{(e_i/\rho)^2+q}-\frac{A+qB}{q^2}\right]\delta_{ij}
\end{equation}
where $e_i$ is the associated eigenvalue of $\langle m|y|n\rangle$. 

\section{Results and discussion}

In this section, we present some illustrative calculations of the energy spectrum and the wavefunction. First, we considered the potential parameters $A=2.0$, $B=\{-12,-14\}$, $C=1.0$, and $q=0.2$ where we summarize the results of the energy spectrum in Table.\ref{tab:2}. We observe a good matching between the results obtained via the numerical Hamiltonian diagonalization (NHD), the potential parameter spectrum (PPS) method, and the asymptotic iteration method (AIM) (Ref.\onlinecite{Sous}).

On the other hand, we considered a larger spectrum size by taking the potential parameters as $A=2.0$, $B=-10.0$, $C=1.0$, and $q=0.2$ which has eight bound states as shown in Table.\ref{tab:3}. As we can see, the results obtained by NHD and the AIM are in good agreement, while errors in the PPS results increase for higher states. This issue arises from the fact that the basis we have chosen in Eq. \ref{eqn:basisTRA} is finite with the constraint $N<-(\mu+\nu+1)/2$ with $\mu$ being energy dependent. This limits the number of data points that the PPS needs to give a better fitting for the higher excited states. Had we known the analytic properties of the orthogonal polynomials $\bar{H}_n^{(\mu,\nu)}(-F^{-1};\ell,\pi/2)$ introduced earlier, we would have been able to obtain exact results. Alternative procedures such as NHD and AIM are necessary to accurately study the higher excited states.

\begin{figure*}
    \centering
    \captionsetup{justification   = raggedright,
              singlelinecheck = false}
    \includegraphics[width=150mm]{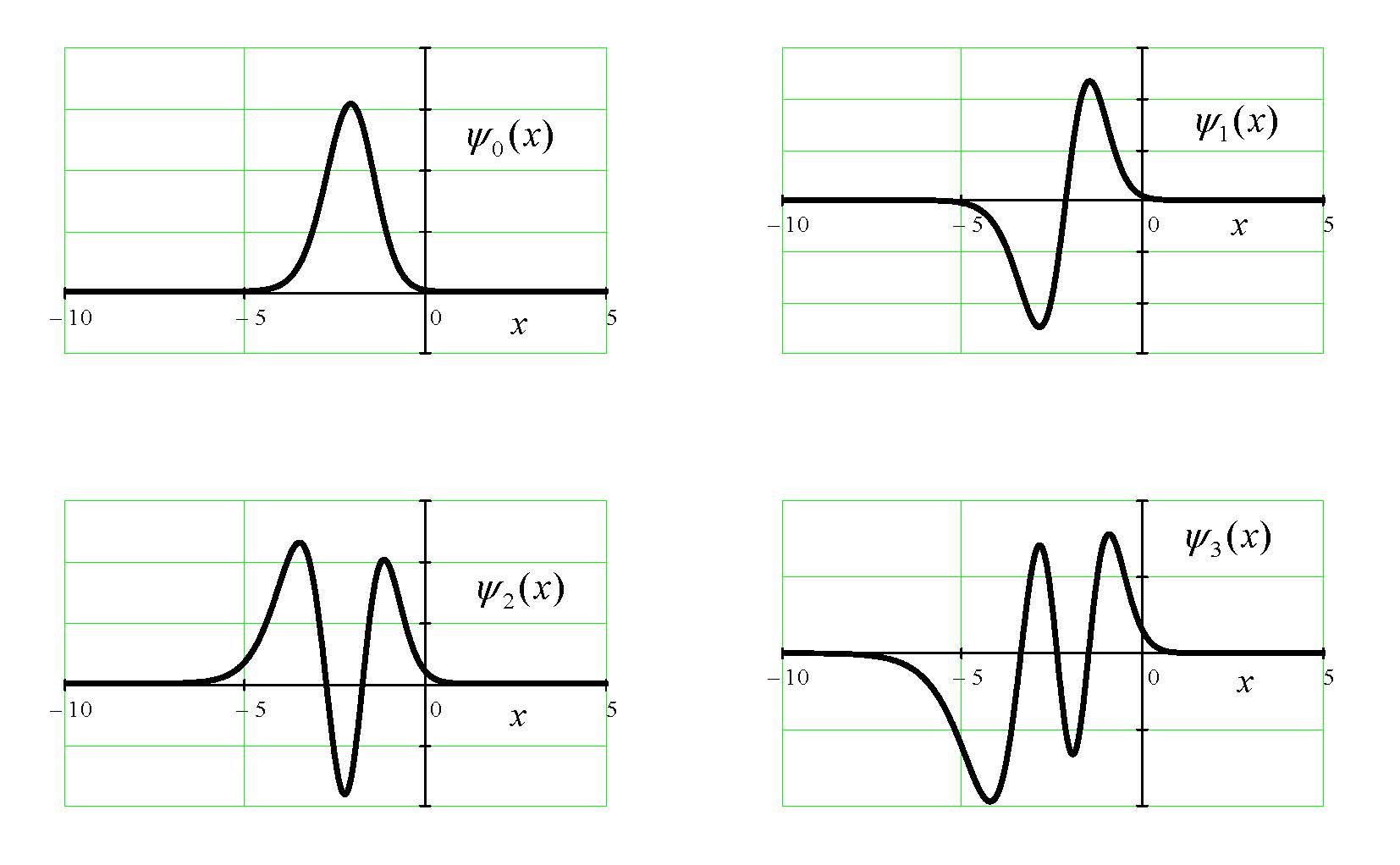}
    \caption{The lowest four bound states of the potential (\ref{eqn:potentialVq}) as generated by the TRA wavefunction (14) for the parameter choices  $A=2.0$, $B=-12$, $C=1.0$, and $q=0.20$.}
    \label{fig:WF}
\end{figure*}

Aside from the energy spectrum, we also plotted the lowest four eigenstates of the potential (\ref{eqn:potentialVq}), for the parameters choice $B =-12$ in Table. \ref{tab:2}, as shown in Fig. \ref{fig:WF}.  

\begin{table*}[ht]
\caption{The complete bound state energy spectrum (in atomic units and with an overall negative sign) for the potential (\ref{eqn:potentialVq})
obtained using the potential parameter spectrum (PPS) technique. For comparison, we present a numerical
Hamiltonian diagonalization (NHD) in the Laguerre basis of size 200 with the Laguerre polynomial index $\gamma$ chosen from the plateau of stability as indicated along with results from the asymptotic iteration method (AIM) with 360 iterations and seed point $x_0$ chosen to be at the local minimum of our potential.\cite{Sous} The potential parameters were chosen as $\{A,C,q,\lambda\}=\{2.0,1.0,0.20,1.0\}$ and we varied the parameter $B$ as shown. }
\begin{center}
\begin{tabular}{p{0.15\linewidth}p{0.15\linewidth}p{0.15\linewidth}p{0.15\linewidth}p{0.15\linewidth}p{0.15\linewidth}}
\hline
\multicolumn{2}{c}{$B=-12$}& &\multicolumn{2}{c}{$B=-14$}\\ 
\hline
NHD ($\gamma=2.0$) & PPS & AIM & NHD ($\gamma=3.0$) & PPS & AIM\\
\hline
6.725966329&6.725966329&6.725966329&3.438724142&3.438724142&3.438724142\\
4.602821791&4.602821791&4.602821791&1.746928179&1.746928179&1.746928106\\
2.795104002&2.795104002&2.795104007&0.550245228&0.550245231&0.5501964482\\
1.348987620&1.348987619&1.348983460&0.008709768&0.008711308&0.026629576\\
0.354453319&0.354453301&0.3556443488&\\
\hline
\end{tabular}
\end{center}
\label{tab:2}
\end{table*}

\begin{table}
\caption{Reproduction of Table \ref{tab:2} with the same parameters except for $B=-10.0$.}
\begin{center}
\begin{tabular}{p{0.3\linewidth}p{0.3\linewidth}p{0.3\linewidth}}
\hline
NHD ($\gamma=1.0$) & PPS & AIM\\
\hline
11.092470042&11.092470042&11.092470042\\
8.789641222&8.789641170&8.789641222\\
6.721176457&6.721163320&6.721176456\\
4.883397609&4.883236520&4.883397612\\
3.275961756&3.275108205&3.275961745\\
1.909788824&1.882238563&1.909788706\\
0.824167396&0.757075573&0.8240517209\\
0.126148628&0.047743887&0.1463293492\\
\hline
\end{tabular}
\end{center}
\label{tab:3}
\end{table}

\section{Conclusions}

In this work, we have studied an interesting and exactly solvable 1D potential (\ref{eqn:potentialVq}) which can support both bound and /or resonance states depending on the choice of potential parameters as shown in the spectral phase diagrams Fig. \ref{fig:phaseS}. We obtained the bound state solution of this potential via the tridiagonal representation approach. The wavefunction is expressed as a finite series in terms of the finite Jacobi polynomials with expansion coefficients expressed in terms of the dipole polynomials $\bar{H}^{\mu,\nu}(z^{-1};\ell,\theta)$. Due to the fact that the analytic properties of the later polynomial are still an open problem to be solved, we relied on numerical means to evaluate the energy spectrum. We have used the potential parameter spectrum, the numerical Hamiltonian diagonalization, and the asymptotic iteration method as three independent techniques and obtained the eigenvalues of our problem for several choices of the potential parameters. The results obtained by the three methods are in good agreement except for the PPS when the size of spectrum gets larger. On the other hand, resonance states can be studied using other well-known methods such as complex scaling. 

Finally, potential (\ref{eqn:potentialVq}) can be used to model diatomic molecules and interactions between atoms and surfaces. It also adds more freedom when compared to the well-known 1D Morse potential due to the fact that the fitting parameters are more than the ones appearing in the Morse potential, and thus being more practical when considering fitting experimental data such as spectroscopic data.   

\begin{acknowledgments}
The authors would like to thank Dr. A. J. Sous, Al-Quds open university, Palestine, for calculating the energy spectrum via the asymptotic iteration method as given in Tables \ref{tab:2} \& \ref{tab:3}. 
\end{acknowledgments}

\appendix

\section{Reviewing the TRA}
\label{eqn:TRA}

The TRA is one of the methods to solve the wave equation by expressing its solutions in terms of orthogonal polynomial. The basic idea of the method is that starting from Schrodinger equation $\left(\mathcal{H}-E\right)\ket{\psi}=0$, we expand the solution as an infinite (or finite) bounded series $\ket{\psi}=\sum\limits_{n}{{f}_{n}}\left( E \right)\ket{\phi_n}$, where $\{f_n(E)\}$ are expansion coefficients which are generally functions of energy and potential parameters, and $\{\ket{\phi_n}\}$ are square integrable basis states. The basis elements usually have the following form 
\begin{equation}
    \langle x|\phi_n\rangle=A_n w(x) P_n(x)
\end{equation}
where $A_n$ is a normalization constant, $w(x)$ is the weight function that vanishes on the boundary of the configuration space, and $P_n(x)$ is an orthogonal polynomial. For example, when $x\geq 0$, one could use Laguerre basis in which $w(x)=x^{\alpha}e^{\beta x}$ and $P_n(x)=L_n^{\nu}(x)$, for some parameters $\{\alpha,\beta,\nu\}$ with $\nu>-1$.

Requiring that the matrix representation of the wave operator $\mathcal{H}-E$ to be tridiagonal in the chosen basis $\{\ket{\phi_n}\}$, the wave equation becomes a three-term recursion relation for the expansion coefficients, 
\begin{equation}
    a_nf_n+b_nf_{n-1}+b_{n+1}f_{n+1}=0
\end{equation}

For some cases, it was found that those expansion coefficients can be related to well-known orthogonal polynomials, while in other situations a new family of orthogonal polynomials arise which requires a further study which is beyond the context of this work.\cite{newpolys} Having the analytic properties of those polynomials will help us find the analytic properties of the system such as bound state energies and the phase shift. In our situation, we have come across new family of orthogonal polynomials that where introduced by Alhaidari in Ref.\onlinecite{newpolys}, and since we still do not know their analytic properties we had to switch to numerical means using the PPS method within the context of the TRA.

\section{Properties of the finite Jacobi polynomials on the semi-infinite interval}
\label{app:Jacobi}
These Jacobi polynomials $P_{n}^{(\mu,\nu)}(x)$ with $x\geq 1$ satisfy the following second order differential equation 
\begin{eqnarray}
\label{eqn:DffJac}
(x^2-1)\frac{d^2P_n^{(\mu, \nu)}(x)}{d x^2}=[\nu-\mu -(\mu + \nu + 2)x]\times
    \nonumber \\
\frac{dP_n^{(\mu, \nu)}(x)}{d x}+ n(n+\mu+\nu+1)P_n^{(\mu, \nu)}(x)&&
\end{eqnarray}
where $n=0,1,\cdots,N$, $\mu>-1$, and $\mu+\nu<-2N-1$. These polynomials satisfy the following properties
\begin{eqnarray}
\lefteqn{\left(x+Q_n\right)P_n^{(\mu, \nu)}(x) = \tfrac{2(n+\mu)(n + \nu)}{(2n
+ \mu + \nu)(2n + \mu + \nu + 1)}\times}
\nonumber\\
&&{} P_{n-1}^{(\mu, \nu)} (x)+ \tfrac{2(n+1)(n + \mu + \nu + 1)}{(2n + \mu + \nu + 1)(2n + \mu + \nu + 2)}P_{n+1}^{(\mu, \nu)}(x)
\end{eqnarray}
\begin{equation}
    \int\limits_{1}^{\infty }dx{{{\left( x-1 \right)}^{\mu }}{{\left( x+1 \right)}^{\nu }}P_{k}^{(\mu ,\,\nu )}(x)}P_{l}^{(\mu ,\,\nu )}(x)=A_{k}^{-2}{{\delta }_{kl}}
\end{equation}
where,
\begin{multline}
\label{eqn:An}
{{A}_{k}}=\sqrt{\frac{(2k+\mu +\nu +1)\Gamma (k+1)\Gamma (k+\mu +\nu +1)}{{{2}^{\mu +\nu +1}}\Gamma (k+\nu +1)\Gamma (k+\mu +1)}}\times\\
\sqrt{\frac{\sin \pi \left( \mu +\nu +1 \right)}{\sin \pi \nu }}
\end{multline}
Finally, 
\begin{equation}
\label{eqn:overlap}
    \langle n|y|m\rangle=-Q_{n}\delta_{nm}+S_{n}\delta_{n,m-1}+S_{n-1}\delta_{n,m+1}
\end{equation}
where 
\begin{equation}
\label{eqn:C}
    Q_{n}=\frac{\mu^{2}-\nu^{2}}{(2n+\mu+\nu)(2n+\mu+\nu+2)}
\end{equation}

\begin{multline}
\label{eqnD}
S_{n}=\frac{2}{2n+\mu+\nu+2}\times\\\sqrt{\frac{(n+1)(n+\mu+1)(n+\nu+1)(n+\mu+\nu+1)}{(2n+\mu+\nu+1)(2n+\mu+\nu+3)}}
\end{multline}
These polynomials satisfy other relations but we listed here the ones that are relevant to this work, for more details see for example Ref.\onlinecite{unconvJacobi}.
\section*{Data Availability Statement}
Data sharing is not applicable to this article as no new data were created or analyzed in this study.

\nocite{*}
\bibliography{Hyperbolic}

\end{document}